\documentclass[%
 reprint,%
 amssymb,amsmath,%
 aip,cha,%
]{revtex4-1}
\pdfoutput=1
\usepackage{float}
\usepackage{hyperref} 
\usepackage{epsfig}
\usepackage[mathscr]{euscript}
\usepackage{amsmath}
\usepackage{graphicx}       
\usepackage{wrapfig}
\usepackage{dcolumn}  
\usepackage{bm}     
\usepackage{amssymb} 
\usepackage{titlesec}
\usepackage{mathtools}
\usepackage{relsize}
\usepackage{cleveref}
\usepackage{comment}
\usepackage{epsf}
\usepackage{amsmath}
\usepackage{amssymb}
\usepackage{esvect}
\usepackage[running]{lineno}
\usepackage{color} 
\DeclareSymbolFont{boldoperators}{OT1}{cmr}{bx}{n}
\SetSymbolFont{boldoperators}{bold}{OT1}{cmr}{bx}{n}
\edef\bar{\unexpanded{\protect\mathaccentV{bar}}\number\symboldoperators16}
\newcommand*\patchAmsMathEnvironmentForLineno[1]{%
  \expandafter\let\csname old#1\expandafter\endcsname\csname #1\endcsname
  \expandafter\let\csname oldend#1\expandafter\endcsname\csname end#1\endcsname
  \renewenvironment{#1}%
     {\linenomath\csname old#1\endcsname}%
     {\csname oldend#1\endcsname\endlinenomath}}%
\newcommand*\patchBothAmsMathEnvironmentsForLineno[1]{%
  \patchAmsMathEnvironmentForLineno{#1}%
  \patchAmsMathEnvironmentForLineno{#1*}}%
\AtBeginDocument{%
\patchBothAmsMathEnvironmentsForLineno{equation}%
\patchBothAmsMathEnvironmentsForLineno{align}%
\patchBothAmsMathEnvironmentsForLineno{flalign}%
\patchBothAmsMathEnvironmentsForLineno{alignat}%
\patchBothAmsMathEnvironmentsForLineno{gather}%
\patchBothAmsMathEnvironmentsForLineno{multline}%
}

\begin{document}
\title{Rare slips in fluctuating synchronized oscillator networks}
\author{Jason Hindes}
\author{Ira B. Schwartz}
\affiliation{U.S. Naval Research Laboratory, Code 6792, Plasma Physics Division, Washington, D.C. 20375, USA}

\begin{abstract}
We study rare phase slips due to noise in synchronized Kuramoto oscillator networks. In the small-noise limit, we demonstrate that slips occur via large fluctuations to saddle phase-locked states. For tree topologies, slips appear between subgraphs that become disconnected at a saddle-node bifurcation, where phase-locked states lose stability generically. This pattern is demonstrated for sparse networks with several examples. Scaling laws are derived and compared for different tree topologies. On the other hand, for dense networks slips occur between oscillators on the edges of the frequency distribution. If the distribution is discrete, the probability-exponent for large fluctuations to occur scales linearly with the system size. However, if the distribution is continuous, the probability is a constant in the large network limit, as individual oscillators fluctuate to saddles while all others remain fixed. In the latter case, the network's coherence is approximately preserved. 

\end{abstract}

\maketitle
{\quotation{Network dynamics is a very
active field of research, and in particular, the study of 
coupled oscillator synchronization. An area of great interest concerns how topology, dynamics, and uncertainty 
conspire to produce rare and extreme events in networked systems. 
In this work, we develop such a theory for the sudden build-up of phase separation in synchronized
oscillator networks, known as slips. Using our theoretical framework,
we discover the underlying mechanisms for the occurrence of slips by finding
their dynamical paths of escape from synchronized states. In particular,
we show how the most probable escape paths and probabilities of occurrence
vary widely as a function of topology and oscillator heterogeneity.}}

\section{\label{sec:Intro} INTRODUCTION}
There is great interest in the formation of spontaneous rhythms in networks of interacting dynamical systems, broadly called synchronization \cite{Arenas}. Synchronization can come in a variety of forms from cluster-synchronization of chaotic systems\cite{Pecora,Sorrentino}, to collective oscillations in coupled limit-cycle oscillators \cite{Kuramoto,Kurths}, and chimera states in spatially extended oscillator networks \cite{Abrams1}. Important practical examples include coupled lasers \cite{Nair}, genetic clocks \cite{Hasty}, and power systems \cite{Timme}. Moreover, many oscillator networks can be approximated by phase-only models, for instance in the limit of weakly coupled limit-cycle oscillators \cite{Kuramoto, Acebron, Strogatz1,Ott}, and deviations from synchronized states \cite{Nishikawa}. Phase-only descriptions have been useful in understanding synchronization in power-grids\cite{DorflerBullo,Nishikawa}, laser arrays \cite{Hoppensteadt}, coupled Josephson junctions \cite{Wiesenfeld}, and functional brain networks \cite{Lee}. Despite their simplicity, such approximations can display a rich variety of synchronization patterns and transitions \cite{Acebron, BulloRev,Hindes1}.

Since oscillator networks are ubiquitous in many real-world settings where noise and uncertainty play a significant role, there is growing interest in understanding the effects of dynamical perturbations and noise on network synchronization \cite{Roberts,Kurths,Kurths2,Pikovsky,Timme2,Kamps,Brezetskiy}. Perhaps the most interesting and important effect of noise in nonlinear network dynamics is the tendency to produce large qualitative changes in the behavior over time, called large fluctuations (LFs). Large fluctuations have been observed in a variety of settings from population extinction \cite{HindesPRL,HindesEPL}, to switching in reaction networks\cite{HindesSR2017,Motter2015PRX}, and power-grid cascades \cite{Nesti}. However, much is yet unknown about how LFs emerge in nonlinear oscillator networks\cite{Timme2,Bouchet}.

The paper layout is the following: In Sec.\ref{sec:SynchAndNoise} we discuss how locked states lose stability at saddle-node bifurcations in Kuramoto networks generically -- implying the existence of saddle-states. The unstable modes of saddles drive networks to produce phase slips when combined with noise. The structure of the unstable modes is discussed for different topologies in Sec.\ref{sec:Component}. How noise causes networks to fluctuate toward saddles is analyzed in Sec.\ref{sec:LargeDeviation}. Mechanisms and scalings are compared and contrasted for tree and dense network topologies in Sec.\ref{sec:Limiting}. Sec. \ref{sec:RareEvent} compares theory to Monte-Carlo simulations. 
\begin{figure*}[t]
\centering
\includegraphics[scale=0.285]{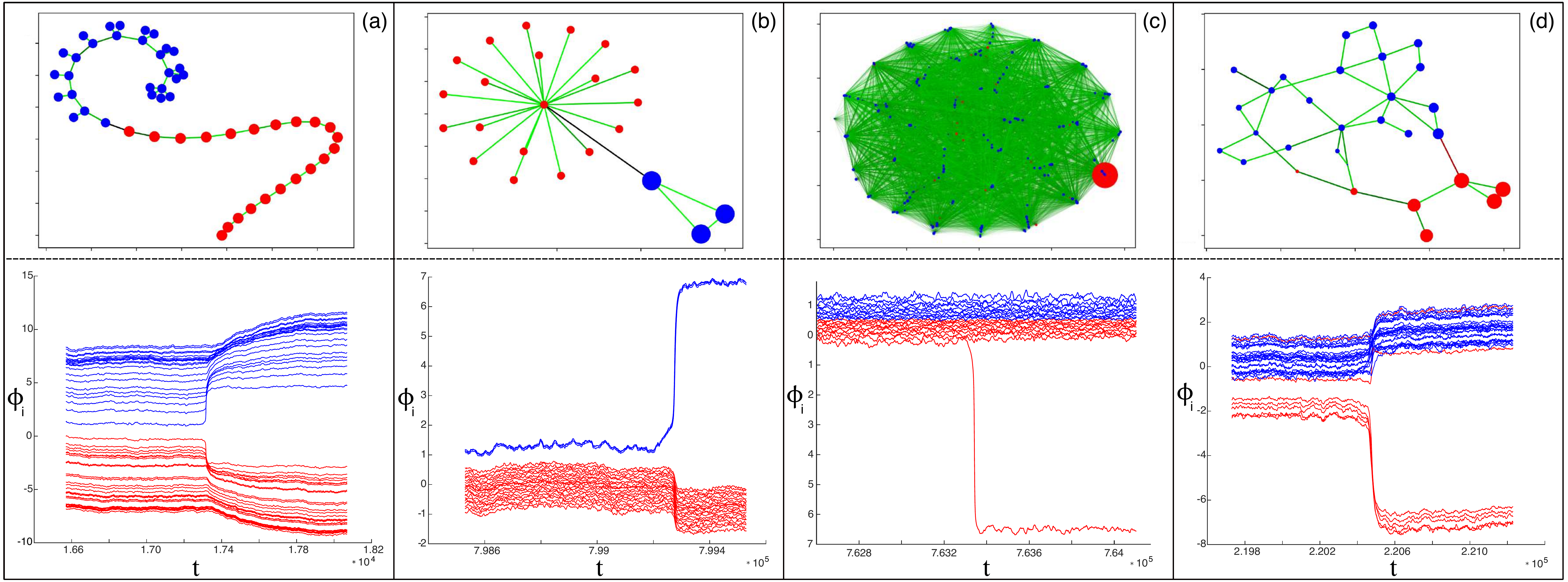}
\caption{Noise-induced rare phase slips. Top panels depict several networks with $J\!\approx\!J_{SN}$. Nodes are colored according to their group in the unstable mode: plus (blue) and minus (red). The sizes of the nodes are proportional to the magnitude of the node's component value, $|{\vec{d}}_{i}^{\;(2)}|$. Edges are drawn in a green scale if $\cos(\phi_{i}^{s}\!-\!\phi_{j}^{s})\!>\!0$ and red scale otherwise, from $|\!\cos(\phi_{i}^{s}\!-\!\phi_{j}^{s})|\!=\!0$ (black) to $|\!\cos(\phi_{i}^{s}\!-\!\phi_{j}^{s})|\!=\!1$ (green). Bottom panels show $\vec{\phi}(t)$ during a slip with $J\!\gtrsim\!J_{SN}$. (a) $J\!=\!0.820$, $D\!=\!0.001$, $T_{c}\!=\!2.0$, $J_{SN}\!=\!0.80405$.  (b) $J\!=\!0.718$, $D\!=\!0.001$, $T_{c}\!=\!2.0$, and $J_{SN}\!=\!0.6957$. (c) $J\!=\!0.64878$, $D\!=\!0.001$, $T_{c}\!=\!2.0$, and $J_{SN}\!=\!0.63818$. (d) $J\!=\!0.322$, $D\!=\!0.001$, $T_{c}\!=\!2.0$, and $J_{SN}\!=\!0.3097$. \cite{F3}} 
\label{StructureAndSlips}
\end{figure*} 
 
\section{\label{sec:SynchAndNoise}PHASE-LOCKED SYNCHRONIZATION IN NETWORKS WITH SMALL NOISE}
Let us consider the Kuramoto model (KM) of synchronization in networked oscillators. The interaction network is defined by a fixed set of (N) nodes and connections between them, 
such that it can be represented by an adjacency matrix, $\boldsymbol{A}$, whose elements are binary, i.e., $A_{ij}\!=\!1$ if nodes i and j are connected and zero otherwise. 
Each node has a phase, $\theta_{i}$, and a natural frequency $\omega_{i}$. The frequencies are fixed in time but heterogeneous. The oscillators have three tendencies in the KM: to oscillate at the natural frequency, to align with neighbors, and 
to fluctuate with random external forces. The dynamics are
\begin{align}
\label{eq:KuramotoWithNoise}
\dot{\theta}_{i}=\omega_{i} + J\sum_{j}A_{ij}\sin(\theta_{j}-\theta_{i}) +\xi_{i}(t),    
\end{align}
where $J$ is a coupling constant. We assume that the random external forces $\xi_{i}(t)$, or {\it noise}, are temporally correlated, but topologically uncorrelated, and are 
given by Ornstein-Uhlenbeck processes with time-correlations $\left<\xi_{i}(t)\xi_{j}(t')\right>\!=\!D\delta_{ij}\exp{\!\{|t-t'|/T_{c}\}}/T_{c}$ $\forall\{i,j\}
\;$, where $T_{c}$ is the correlation time, and D is the noise intensity. Temporal correlations are known to have a significant effect on noise-induced desynchronization in power grids \cite{Kamps}.   

In this work we study the effects of small noise on phase-locked synchronized states of Eq.(\ref{eq:KuramotoWithNoise}). As we will see, such noise produces
large and rare fluctuations with a particular structure that depends on the network topology and natural frequencies. First, it is useful to discuss synchronized states and their generic bifurcations 
when $D\!=\!0$ before considering noise.

In a state of phase-locked synchronization (PL) all nodes oscillate at the average frequency, $\left<\omega\right>\!\equiv\!\sum_{i}\omega_{i}/N$, or $\theta_{i}(t)\!=\!\phi_{i}^{*}+\left<\omega\right>\!t$, where 
$\phi_{i}^{*}$ satisfies the equations: 
\begin{linenomath} 
\begin{align}
\label{eq:Synchronization}
0=\omega_{i}-\left<\omega\right> + J\sum_{j}A_{ij}\sin(\phi_{j}^{*}-\phi_{i}^{*}) \;\;\;\;\forall i.     
\end{align}
\end{linenomath} 
Solutions of Eq.(\ref{eq:Synchronization}) are locally stable if $J$ is sufficiently large\cite{Manik,BulloRev}. Local stability is determined by the eigenvalues of the Jacobian matrix for $\dot{\theta}_{i}$, whose elements are 
\begin{align}
\label{eq:SynchronizationJacobian}
\tilde{L}_{ij}(\vec{\phi}^{*})=J\Big[A_{ij}\cos(\phi_{i}^{*}-\phi_{j}^{*})-\delta_{ij}\!\sum_{k}A_{ik}\cos(\phi_{i}^{*}-\phi_{k}^{*})\!\Big]\!. 
\end{align}
It is important to note that Eq.(\ref{eq:SynchronizationJacobian}) is proportional to the {\it Laplacian} of the symmetric weighted network, $\tilde{\boldsymbol{A}}(\vec{\phi}^{*})$, 
\begin{align}
\label{eq:AlignmentNetwork}
\tilde{A}_{ij}(\vec{\phi}^{*})=A_{ij}\cos(\phi_{i}^{*}\!-\!\phi_{j}^{*})
\end{align}
-- namely the interaction network with weights given by the phase alignment between pairs of connected nodes\cite{BulloRev}. We note that in networks with special symmetries, such as cyclical topology and homogeneous frequencies, there can exist stability between multiple PL states\cite{Timme3}. In this work, we focus on a single stable $\vec{\phi}^{*}$ (modulo $2\pi$).  

The Laplacian form of the Jacobian has several implications. First, because every row sums to zero, there is an eigen-solution (or mode) with zero eigenvalue, $\vec{0}=\boldsymbol{\tilde{L}}(\vec{\phi}^{*}){\vec{r}}^{\;(1)}$, where all nodes (or components) have equal values, $r_{i}^{(1)}\!=\!r$. This neutral mode is a consequence of rotational symmetry in Eq.(\ref{eq:KuramotoWithNoise}). Since the average phase is conserved in a frame rotating with the average frequency, $\vec{\phi}\!=\!\vec{\theta}\!-\!\left<\omega\right>\!t$, a single oscillator $l$ can be set to $\phi_{l}\!=\!N\!\left<\phi\right>\!-\!\sum_{j\neq l}\!\phi_{j}$; in which case, the neutral mode is removed. If all other eigenvalues of $\boldsymbol{\tilde{L}}(\vec{\phi}^{*})$ are negative, $\vec{\phi}^{*}$ is locally stable. This is easy to see if $\tilde{\bold{A}}(\vec{\phi}^{*})$ is non-negative and symmetric, since it follows that $0\!\geq\!p^{(2)}\!\geq\!p^{(3)}\!\geq\!...\!\geq\!p^{(N)}\;\forall\; l\!>\!1$ with $p^{(l)}{\vec{r}}^{\;(l)}\!=\!\boldsymbol{\tilde{L}}(\vec{\phi}^{*}){\vec{r}}^{\;(l)}$ \cite{BulloRev,Newman3}. Second, all other modes with negative eigenvalues sum to zero $\sum_{i}r_{i}^{(l)}\!=\!0\;\forall\; l>1$ \cite{SM}. This second property implies that each non-neutral mode can be separated into two groups of nodes for which $r_{i}^{(l)}\!>\!0$ and $r_{i}^{(l)}\!<\!0$. Third, because $\boldsymbol{\tilde{L}}(\vec{\phi}^{*})$ is real and symmetric, all eigenvalues are real. 

Stability of $\vec{\phi}^{*}$ is lost as we decrease $J$ from infinity. A single eigenvalue crosses zero , $p^{(2)}\!=\!0$, at a co-dimension-one {\it saddle-node bifurcation} (SN), as shown in \cite{Manik,Mirollo,Pazo}. At this point, $\vec{\phi}^{*}$ collides with a saddle and disappears. We denote the coupling at bifurcation $J_{SN}$,
\begin{align}
\label{eq:SaddleNode}
p^{(2)}(J\!=\!J_{SN})=0.
\end{align}

Because of the form of SN bifurcations, when $J\!\gtrsim\!J_{SN}$ there exists a {\it saddle} phase-locked state, ${\vec{\phi}}^{s}$, also satisfying Eq.(\ref{eq:Synchronization}) \cite{Kuznetsov1}. In addition, the Jacobian for the saddle has a similar form as $\vec{\phi}^{*}$ with eigenmodes $s^{(l)}{\vec{d}}^{\;(l)}\!=\!\boldsymbol{\tilde{L}}(\vec{\phi}^{\;s}){\vec{d}}^{\;(l)}$. Importantly, however, the eigenvalue for ${\vec{d}}^{\;(2)}$ is non-negative, $s^{(2)}\!\geq\!0$. As a consequence of the second property stated above, ${\vec{\phi}}^{s}$ {\it has at least one unstable mode which tends to separate two groups of oscillators dynamically}. We define these groups by the sets $\mathcal{P}$ and $\mathcal{M}$, {\it plus} and {\it minus}, with $i\in \mathcal{P}$ if ${d_{i}}^{(2)}\!>\!0$ and $i\in \mathcal{M}$ if ${d_{i}}^{(2)}\!<\!0$.
       
\subsection{\label{sec:Component} Unstable modes and slips}

So far the dynamics considered have been deterministic. When a synchronized network is subjected to non-zero but small noise $(D\!\ll\!1)$, ${\vec{\phi}}(t)$ fluctuates in a region around ${\vec{\phi}}^{*}$ for an exponentially long time. Eventually, however, a noise sequence ${\vec{\xi}}(t)$ is generated that carries ${\vec{\phi}}(t)$ close to ${\vec{\phi}}^{s}$. Once the saddle is reached along a fluctuational path, the most-likely subsequent event is for the system to follow a deterministic trajectory and ``roll downhill" along an unstable direction of the saddle. As we have argued above, this dynamics tends to separate the phases of oscillators in $\mathcal{P}$ from oscillators in $\mathcal{M}$. In the most severe cases, nodes in $\mathcal{P}$ undergo phase slips with respect to their neighbors in $\mathcal{M}$. A phase slip denotes a phase difference appearing between two nodes in a network that is greater than $2\pi$. A natural question, then, is how are $\mathcal{P}$ and $\mathcal{M}$ related to the topology and frequency distribution, and which oscillators slip? 

Several examples \cite{F3} are given in Fig.\ref{StructureAndSlips}, showing both the structure of unstable modes for several networks (top), and a rare slip time-series for each (bottom). The top and bottom panels of Fig.\ref{StructureAndSlips} (a) show results for a tree network. We can see that at the SN bifurcation, a single edge has $\cos(\phi_{i}^{s}\!-\!\phi_{j}^{s})\!=\!0$. In fact, this is always the case for connected trees, for which the removal of a single edge disconnects the network into two components. We can demonstrate this property through the following argument: as $J$ is decreased toward $J_{SN}$, the alignment between nodes in $\vec{\phi}^{*}$ decreases and $\tilde{A}_{ij}(\vec{\phi}^{*})$ decreases for $A_{ij}\!\neq\!0$. In the absence of special symmetries, a particular edge between two nodes $k$ and $l$ approaches $\tilde{A}_{kl}\!\rightarrow\!0$ first, while $\tilde{A}_{ij}\!>\!0$ $\forall \{i,j\}\!\neq\!\{k,l\}\;\text{and}\;A_{ij}\!\neq\!0$. The edge is effectively removed from the tree and disconnects it into two components by definition \cite{Dekker}. Since $\tilde{\boldsymbol{A}}(J\!=\!J_{SN})$ is non-negative, its Laplacian has exactly two zero eigenvalues, $p^{(2)}\!=\!p^{(1)}\!=\!0$ \cite{Newman3} and the SN condition Eq.(\ref{eq:SaddleNode}) is satisfied. Therefore in the special case of trees, $\mathcal{P}$ and $\mathcal{M}$ are equal to the two disconnected subgraphs at the SN bifurcation, and ${d_{i}}^{(2)}\!=\!\sqrt{|\mathcal{M}|/{|\mathcal{P}|N}}\;\forall i\!\in\!\mathcal{P}$ and ${d_{i}}^{(2)}\!=-\!\sqrt{|\mathcal{P}|/{|\mathcal{M}|N}}\;\forall i\!\in\!\mathcal{M}$ \cite{SM}. Note that nodes within each set are drawn with identical sizes in Fig.\ref{StructureAndSlips} (a) top. As expected, nodes in $\mathcal{P}$ and $\mathcal{M}$ undergo phase-slips with respect to each other, as shown in Fig.\ref{StructureAndSlips}. Large fluctuations in trees are analyzed in Sec.\ref{sec:Limiting}. 

More generally, the subgraph structure of $\mathcal{P}$ and $\mathcal{M}$ is approximately maintained for tree-like and sparse networks, where subsets within each have roughly equal ${d_{i}}^{(2)}$, and slip together. Sparse topology is observed in such relevant technological networks as power grids\cite{Kurths2}. Examples are shown in Fig.\ref{StructureAndSlips} for a star-like network in (b), and an IEEE test-bus network in (d). In the IEEE example, we can see that $\mathcal{M}$ is composed of seven nodes -- five of which have approximately equal $|{d_{i}}^{(2)}|$ and form a subgraph. Likewise, the latter five nodes slip with respect to their neighbors outside of the subgraph during a LF, as shown in the bottom panel. 

Conversely, for large dense networks the unstable modes are generally localized around nodes on the edges of the frequency distribution. An example is shown in Fig.\ref{StructureAndSlips} (c), where $\mathcal{M}$ is composed, effectively, of a single node with the minimum frequency, whereas $\mathcal{P}$ is composed of many nodes with ${d_{i}}^{(2)}\!\sim\!\mathcal{O}(1/N)$. Properties of LFs in dense networks are discussed in Sec.\ref{sec:Limiting}.   

For arbitrary networks and frequencies it is difficult to predict which nodes slip during a LF in general. However, this question can be answered from a computational perspective by computing the (deterministic) unstable eigenvectors of the saddle, placing an initial condition on an eigenvector near the saddle, and running the deterministic equations such that the network evolves back to $\vec{\phi}^{*}$, but with phase-slips.   
Such a trajectory is a heteroclinic connection of Eq.(\ref{eq:KuramotoWithNoise}), which we call H2. Nodes that slip along H2 are nodes that slip in a LF, after noise has driven the network to a saddle. 

\section{\label{sec:LargeDeviation}LARGE--FLUCTUATION PICTURE}
The discussion so far has concerned dynamics of oscillator networks near a saddle point. However, we would like to understand the stochastic process by which noise drives a network from ${\vec{\phi}}^{*}$ to ${\vec{\phi}}^{s}$, predict the probability of reaching ${\vec{\phi}}^{s}$, etc. In the limit of small noise amplitude, the LFs shown in Fig.\ref{StructureAndSlips} are {\it rare events} -- appearing on time scales much longer than the deterministic dynamics. A general feature of such events is that they occur with a probability that is exponential in $1/D$, and arise through an optimal noise sequence that is exponentially more probable than all others\cite{Maier}. This is a well known prediction of large-deviation (fluctuation) theory \cite{Friedlin,Mark1,Schuss,Eric}. The optimal-noise realization of interest is one in which work is done on the system, effectively driving the network from ${\vec{\phi}}^{*}$ to ${\vec{\phi}}^{s}$. Such optimal noise is describable in terms of analytical mechanics in the following way. 

Since the noise is generated from independent Ornstein-Uhlenbeck processes for each node, we have 
\begin{align}
\dot{\xi}_{i}=-\frac{\xi_{i}}{T_{c}} + \frac{\sqrt{2D}}{T_{c}}\eta_{i}(t), 
\label{OrsteinUhlenbeck}     
\end{align}
where $\eta_{i}$ is a zero-mean, Gaussian white-noise source for each node with unit variance. The statistical weight for a given realization of $\vec{\eta}(t)$ scales as\cite{Feynman} 
\begin{align}
\rho[\vec{\eta}(t)]\sim\exp\Big\{\!-\!\sum_{i}\!\int\frac{\eta_{i}^{2}}{2}dt\Big\}. 
\label{Probability}     
\end{align}
The probability exponent for a noise realization is called the {\it action}, $\mathcal{S}$, with $\rho[\vec{\eta}(t)]\!\sim\!\exp{\!\{-\mathcal{S}\}}$. The optimal (most-likely) LFs, therefore, should minimize the action subject to the equality constraints Eq.(\ref{eq:KuramotoWithNoise}) and Eq.(\ref{OrsteinUhlenbeck}). A local minimization of the action can be performed by introducing Lagrange multipliers \cite{Smelly}, $\vec{\lambda}$: 
\begin{align}
\mathcal{S}(\vec{\theta},\vec{\lambda},\vec{\xi},\dot{\vec{\xi}}\;)=&\sum_{i}\!\int\!\Bigg[\!\frac{(\dot{\xi}_{i}T_{c}+\xi_{i})^{2}}{4D}+\frac{\lambda_{i}}{2D}\Big(\dot{\theta}_{i}-\omega_{i}\nonumber\\
&-J\sum_{j}A_{ij}\sin(\theta_{j}-\theta_{i})-\xi_{i}\Big)\!\Bigg]dt. 
\label{Action}     
\end{align}
Hence, the {\it optimal noise producing LFs satisfies Euler-Lagrange equations}: 
\begin{align}
\label{Lagrange1} 
\dot{\phi}_{i}&=\omega_{i}-\left<\omega\right> + J\sum_{j}A_{ij}\sin(\phi_{j}-\phi_{i}) +\xi_{i},\\ 
\label{Lagrange2} 
\dot{\lambda}_{i}&=-\sum_{j}\tilde{L}_{ij}(\vec{\phi})\lambda_{j},\\
\label{Lagrange3}  
T_{c}^{2}\ddot{\xi}_{i}&=\xi_{i}-\lambda_{i}.       
\end{align}

\begin{figure}
\centering
\includegraphics[scale=0.33]{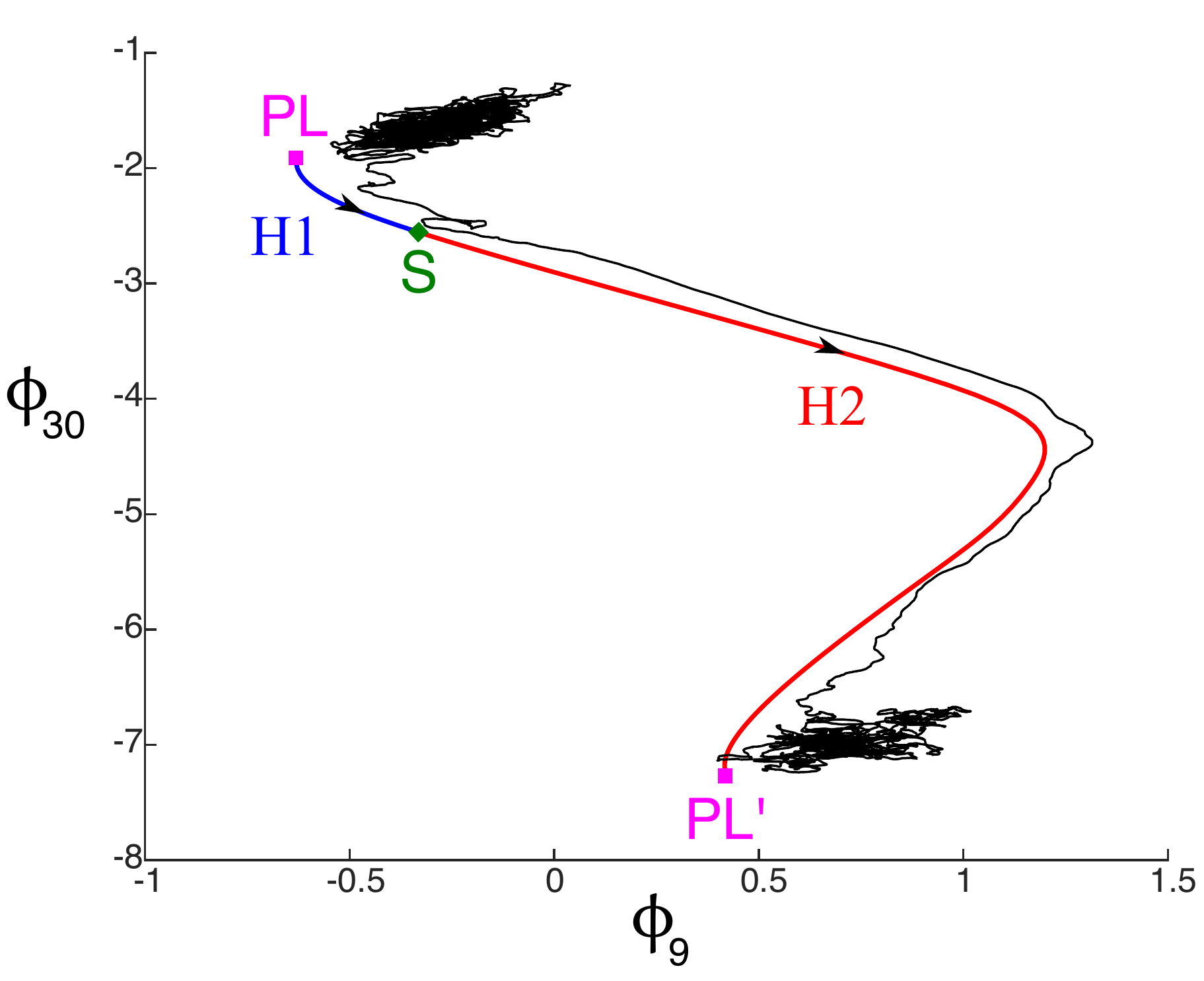}
\caption{Large fluctuation from a phase-locked state $(PL)$ to a phase-locked state with phase-slips $(PL^{'})$ through a two-step switching mechanism: (H1) noise-induced heteroclinic connection along a stable manifold (blue) of the saddle $(S)$. (H2) zero-noise heteroclinic connection along an unstable manifold (red) of the saddle. The paths are projected into the phases of two oscillators for the IEEE network \cite{F3}. A randomly selected stochastic trajectory is shown in black (slightly shifted) for comparison: $J\!=\!0.330$, $T_{c}\!=\!2.0$ and $D\!=\!0.0018$.}
\label{Schematic}
\end{figure}

Of interest for predicting the noise-induced phase slips described in Sec.\ref{sec:Component}, are particular solutions of Eqs.(\ref{Lagrange1}-\ref{Lagrange3}) that start at ${\vec{\phi}}^{*}$ and end at ${\vec{\phi}}^{s}$ $\;$\cite{Maier}. Such solutions define heteroclinic connections, which we call H1, that are distinct from H2. The latter requires no noise. In practice, H1 must be constructed numerically, by solving Eqs.(\ref{Lagrange1}-\ref{Lagrange3}) subject to the boundary conditions: $\vec{\phi}(t\!=\!-\infty)\!=\!{\vec{\phi}}^{*}$, $\vec{\lambda}(t\!=\!\!-\infty)\!=\!\vec{\xi}(t\!=\!\!-\infty)\!=\!\dot{\vec{\xi}}(t\!=\!\!-\infty)\!=\!\vec{0}$, and $\vec{\phi}(t\!\rightarrow\!\infty)\!=\!{\vec{\phi}}^{s}$, $\vec{\lambda}(t\!\rightarrow\!\infty)\!=\!\vec{\xi}(t\!\rightarrow\!\infty)\!=\!\dot{\vec{\xi}}(t\!\rightarrow\!\infty)\!=\!\vec{0}$ $\;$\cite{Schuss}. A schematic is given in Fig.\ref{Schematic}, showing both H1 and H2 for the example IEEE network. We point out that the H1 and H2 sequence is the characteristic form for noise-induced {\it switching} between stable fixed points \cite{Mark2}. 

Before looking at limiting cases, we first consider general features of Eqs.(\ref{Lagrange1}-\ref{Lagrange3}). First, since the integrand in Eq.(\ref{Action}) has no explicit time dependence, the energy  
\begin{align}
\label{Energy} 
E=\frac{1}{2D}\sum_{i}\!\Big[\lambda_{i}\dot{\phi}_{i}+\frac{1}{2}(T_{c}^{2}\dot{\xi_{i}}^{2}\!-\xi_{i}^{2})\Big],       
\end{align}
is conserved. Given the boundary conditions, energy conservation implies that H1 and H2 {\it are zero-energy invariant manifolds}, $E\!=\!0$. 

Second, summing Eq.(\ref{Lagrange2}) over all nodes implies that $\dot{\left<\lambda\right>}\!=\!0$. Because of the boundary conditions for H1 and H2, summing over Eq.(\ref{Lagrange1}) and Eq.(\ref{Lagrange3}) gives 
\begin{align}
\label{Constant} 
\dot{\left<\phi\right>}=\left<\lambda\right>=\left<\xi\right>=0.
\end{align}
The condition Eq.(\ref{Constant}) implies that {\it along a LF from} ${\vec{\phi}}^{*}$, {\it the average phase} (averaged over all nodes) {\it is conserved}. Similarly, the noise on the network averages to zero, and therefore no work is done by the noise against the neutral mode. 

\subsection{\label{sec:Limiting} Limiting cases}
It is useful to consider the white-noise or memoryless limit, $T_{c}\!\rightarrow\!0$, typically assumed in most works on the KM \cite{Acebron,Strogatz1,Kurths}. We note that Eq.(\ref{Lagrange1}), can be written in the form of a gradient system with additive noise, $\dot{\vec{\phi}}\!=\!-\partial U\!/\!\partial{\vec{\phi}}\!+\!\vec{\xi}$, and
\begin{align}
\label{Potential}
U(\vec{\phi})=-\sum_{i}(\omega_{i}\!-\!\left<\omega\right>)\phi_{i} -\frac{J}{2}\sum_{ij}A_{ij}\cos(\phi_{i}-\phi_{j}). 
\end{align}
When $T_{c}\!=\!0$, it is straightforward to show that $\mathcal{S}$ is minimized when $\dot{\vec{\phi}}\!=\!\partial U\!/\!\partial{\vec{\phi}}$, implying that optimal fluctuations are time-reversed relaxations, and therefore H1 is a heteroclinic connection of the noise-free system \cite{Eric, Onsager}. As a consequence of reversibility, the action in the white-noise limit, $\mathcal{S}_{W}$, has a simple interpretation as proportional to the difference in the KM potential function\cite{Timme2}, Eq.(\ref{Potential}),   
\begin{align}
\label{ActionWN}
\mathcal{S}_{W}(\vec{\phi})=\frac{1}{D}[U(\vec{\phi})-U(\vec{\phi}^{*})]. 
\end{align}

For general networks and natural frequencies, Eq.(\ref{ActionWN}) must be solved numerically. However, analytic insight can be gained on the effects of topology on LFs by looking near bifurcation, $J\!=\!J_{SN}(1+\delta)$, where $\delta\!\ll\!1$. For example, in tree networks the critical coupling, $J_{T}\!\equiv\!J_{SN}$, can be calculated exactly by summing Eq.(\ref{eq:Synchronization}) over $\mathcal{P}$ with $\delta\!=\!0$. Since there is only one edge connecting a single oscillator in $\mathcal{P}$ to a single oscillator in $\mathcal{M}$, with a relative phase difference between the oscillators of $\pi/2$,
\begin{align}
\label{TreeCoupling}
J_{T}=\left|\sum_{i\in\mathcal{P}}\!\left<\omega\right>-\omega_{i}\right|,   
\end{align}
as found in \cite{Dekker}. Expanding $\vec{\phi}^{s}-\vec{\phi}^{*}$ in powers of $\delta$ in Eq.(\ref{eq:Synchronization}) and Eq.(\ref{ActionWN}) \cite{SM}, we find $\phi_{i}^{s}\!-\!\phi_{i}^{*}\!\approx\!2\sqrt{2}|\mathcal{M}|\delta^{\frac{1}{2}}\!/N$ if $i\!\in\!\mathcal{P}$, $\phi_{i}^{s}\!-\!\phi_{i}^{*}\!\approx\!-2\sqrt{2}|\mathcal{P}|\delta^{\frac{1}{2}}\!/N$ if $i\!\in\!\mathcal{M}\;$, and  
\begin{align}
\label{TreeActionNear}
S_{W}\approx\frac{4\sqrt{2}}{3}\delta^{\frac{3}{2}}J_{T} , 
\end{align}
where $|\mathcal{P}|$ is the number of nodes in $\mathcal{P}$. The $3/2$ exponent in Eq.(\ref{TreeActionNear}) is standard for SN bifurcations in homogeneous and well-mixed systems\cite{Ira}. 

\begin{figure}[h]
 \centering
\includegraphics[scale=0.40]{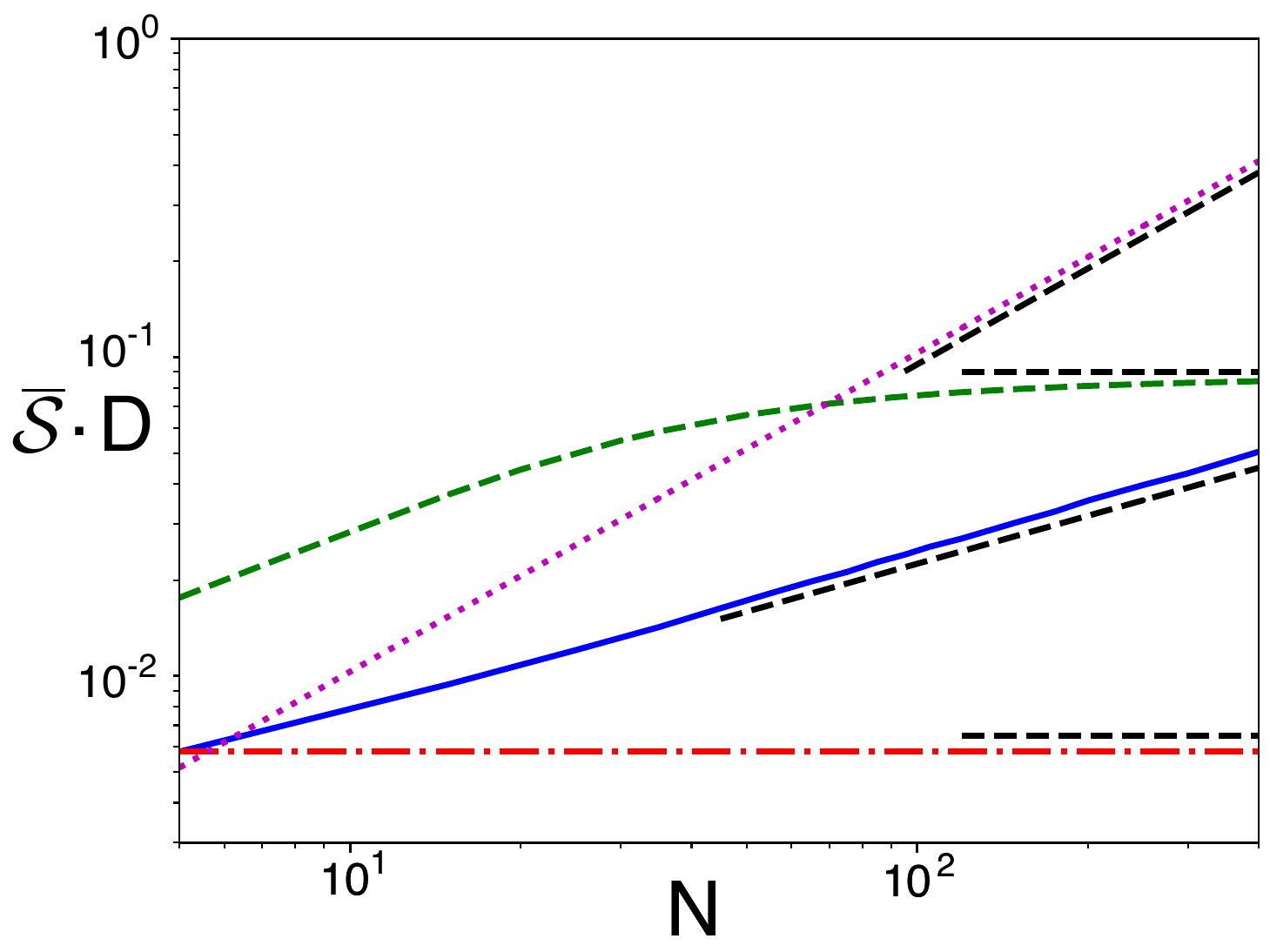}
\caption{Expected action versus $N$ in the white-noise limit, Eq.(\ref{ActionWN}) for four networks: star (red $-.$) with $\omega_{i}\!=\!-0.5\!+\!(i-1)\!/(N\!-\!1)$ and the central node $i\!=\!N$; linear chain (blue $-$) with $\omega_{i}\!=\!-0.5\!+\!(i-1)\!/(N\!-\!1)$; complete graph (green $--$) with $\omega_{i}\!=\!-0.5\!+\!(i-1)\!/(N\!-\!1)$; complete graph (magenta $:$) with $\omega_{i}\!=\!-0.5$ if $i\!\leq\!N/2$ and $\omega_{i}\!=\!0.5$ otherwise. For the linear chain, the action was averaged over $10^{4}$ random shufflings of the nodes. $J\!=\!1.05\!\cdot\!J_{SN}$ for all networks. Black dashed-lines show the predicted scalings.}
\label{ActionScaling}
\end{figure}

Still, the expression is interesting, since at a constant distance to bifurcation $\delta$, the action for different tree topologies differs only in $J_{T}$. In the white-noise limit of the KM, $J_{T}$ is the so called ``topological factor" for LFs \cite{HindesPRL,HindesSR2017,HindesPRE2017}, and the action scales in different ways for different networks depending on how $J_{T}$ behaves. Of course, even if the network and the fraction of nodes with a given frequency are fixed, $J_{T}$ depends on how the frequencies correlate with the underlying tree. Nevertheless, it is possible to derive scalings for the expectation value of $J_{T}$ in certain cases. For instance, it has been shown in \cite{Dekker}, that in the limit of large $N$ and a uniform distribution of frequencies, the expected value of $J_{T}\!\sim\!\mathcal{O}(\sqrt{N})$ for chain networks and k-regular trees, and $J_{T}\!\sim\!\mathcal{O}(1)$ for star networks. Both scalings are intuitive. For the star network, a single node slips 
during a LF, and therefore the probability should be independent of system size. For the chain network, the scaling follows the expected displacement of a random walk with $\mathcal{O}(N)$ steps. In fact, the chain network scaling gives an upper bound for the expectation value of $J_{T}$ in general\cite{Dekker}. Examples are shown in Fig.\ref{ActionScaling} for two tree topologies, where the two limiting scalings with $N$ are demonstrated for the expected action.  

The properties of LFs in trees can be contrasted with dense networks, such as complete graphs (CGs), where every pair of nodes is directly connected \cite{Strogatz1,Mirollo,Kuramoto}. In order to compare networks with finite interactions as $N\!\rightarrow\!\infty$, usually $J\!\rightarrow\!J/N$ for CGs in Eq.(\ref{eq:KuramotoWithNoise}), and we follow this convention. In such networks, if the frequency distribution is discrete, $\mathcal{S}$ scales linearly with $N$. For example, given a fixed set of frequencies $\Omega\!=\!\{\omega_{1},\omega_{2},...\}$, with $|\Omega|\!\sim\!\mathcal{O}(1)$, and a set of fractions $F\!=\!\{f_{1},f_{2},...\}$, where $f_{1}N$ nodes have frequency $\omega_{1}$, and all $f$ are $\mathcal{O}(1)$, Eq.(\ref{Action}) reduces to $\mathcal{S}\!=\!Ns(\Omega,F,J,T_{c})/4D$, where the function $s$ is independent of $N$ \cite{SM}. The linear dependence of the action with $N$ is typical for LFs in ``well-mixed" discrete networks \cite{HindesSR2017,HindesPRE2017}, since $\mathcal{O}(N)$ nodes fluctuate to a saddle. An example is shown in Fig.\ref{ActionScaling} with a bimodal distribution demonstrating the expected scaling, $\bar{\mathcal{S}}$.   

On the other hand, if the distribution of frequencies is continuous, $f(\omega)$, the scaling of the action with $N$ can be quite different as $N\!\rightarrow\!\infty$. For instance, if $f(\omega)$ is continuous over a finite interval, $-\omega_{m}\!\leq\!\omega\!\leq\!\omega_{m}$, then the stable PL is given by   
\begin{align}
\label{PhiStar}
\phi^{*}(\omega)=\sin^{-1}\!\Big(\frac{\omega}{JR^{*}}\!\Big),    
\end{align}
with $-\pi/2\leq\phi^{*}(\omega)\leq\pi/2$ and 
\begin{align}
\label{Order}
R^{*}=\int_{-\omega_{m}}^{\omega_{m}}\!f(\omega)\sqrt{1-\Big(\frac{\omega}{JR^{*}}\!\Big)^{2}}.  
\end{align}
Here, we have used the standard reduction to a single coherence order-parameter, $R\!=\!\left<e^{i\phi}\right>$, for CGs \cite{Kuramoto, Acebron, Strogatz1,Ott}. Importantly, given a particular value for $R^{*}$, we note that there are many saddle-states with $R\!\approx\!R^{*}$ satisfying Eq.(\ref{eq:Synchronization}), but with individual oscillators reflected over $\phi\!=\!\pi/2$. Let us define an index set, $\mathcal{I}$, specifying which oscillators are reflected in the saddle. Namely, given $\mathcal{I}$ there is a saddle state   
\begin{align}
\label{saddle}
\phi^{s}(\omega_{i})=&\;\pi-\phi^{*}(\omega_{i}),\;\text{if}\; i\in\mathcal{I} \;\text{and} \nonumber \\ 
\phi^{s}(\omega_{i})=&\;\phi^{*}(\omega_{i}),\; \text{otherwise}.
\end{align}
As long as $|\mathcal{I}|$ is $\mathcal{O}(1)$, $R(t)\!=\!R^{*}$ along the unstable manifolds of the saddle Eq.(\ref{saddle}) in the limit $N\!\rightarrow\!\infty$. 

In particular, since $R$ is constant, the reflected oscillators have independent dynamics $\phi_{i}(t)\!=\!\phi^{s}(\omega_{i})\!-\!\epsilon_{i}(t)$ for $i\!\in\!\mathcal{I}$, with $\dot{\epsilon}_{i}\!=\!-\omega_{i}\!+\!JR^{*}\sin\{\phi^{*}(\omega_{i})\!+\!\epsilon_{i}\}\;$ \cite{SM}. Using the reversibility property of the white-noise limit, we can compute the minimum action associated with slips of nodes in $\mathcal{I}$ to $\mathcal{O}(T_{c}^2)$, $\mathcal{S}(\mathcal{I})\!\approx\!\sum_{i\in\mathcal{I}}\int[\dot{\epsilon_{i}}^{2}\!+\!T_{c}^{2}\ddot{\epsilon_{i}}^{2}]dt/\!D$ \cite{SM}:
\begin{align}
\label{EdgeFlip}
&\mathcal{S}(\mathcal{I})D\!\approx\!\sum_{i\in\mathcal{I}}\!\Bigg[\omega_{i}\big(2\phi^{*}\!(\omega_{i})\!-\!\pi\big)\!\!+\!2JR^{*}\!\!\cos\{\phi^{*}\!(\omega_{i})\!\}\!+\!\frac{\!T_{c}^{2}\!J^{2}\!{R^{*}}^{\!2}}{2}\nonumber \\
&*\!\!\bigg(\!\omega_{i}\Big(\!2\phi^{*}\!(\omega_{i})\!-\!\pi\!+\!\sin\{2\phi^{*}\!(\omega_{i})\!\}\!\Big)\!+\!\frac{3JR^{*}}{4}\!\cos^{3}\!\{\phi^{*}\!(\omega_{i})\!\}\!\!\bigg)\!\Bigg]\!. 
\end{align} 
The expression Eq.(\ref{EdgeFlip}) is independent of $N$, since the number of slipped nodes is $\mathcal{O}(1)$ by assumption. Of course, {\it the slip with the minimum action in} Eq.(\ref{EdgeFlip}) {\it corresponds to the slip of a single 
oscillator with the maximum (or minimum) frequency}, as mentioned in Sec.\ref{sec:Component}. An example is shown in Fig.\ref{ActionScaling} for a uniform distribution, which asymptotically approaches Eq.(\ref{EdgeFlip}) as $N\!\rightarrow\!\infty$, given exactly two slipped nodes with the maximum and minimum frequency. 

\subsection{\label{sec:RareEvent} Rare-slip observables and simulations}
Now that we have characterized the LFs of phase-locked states, described their relationship to topology and natural frequencies, and constructed a mechanics for predicting most-likely realizations, let us compare predictions to stochastic simulations \cite{SM}. We note that the frequencies in simulations were chosen randomly from a uniform distribution \cite{F3}. However, the distribution details do not affect the global-dynamical structure of LFs (see Fig.\ref{Schematic}).

\begin{figure}
\includegraphics[scale=0.237]{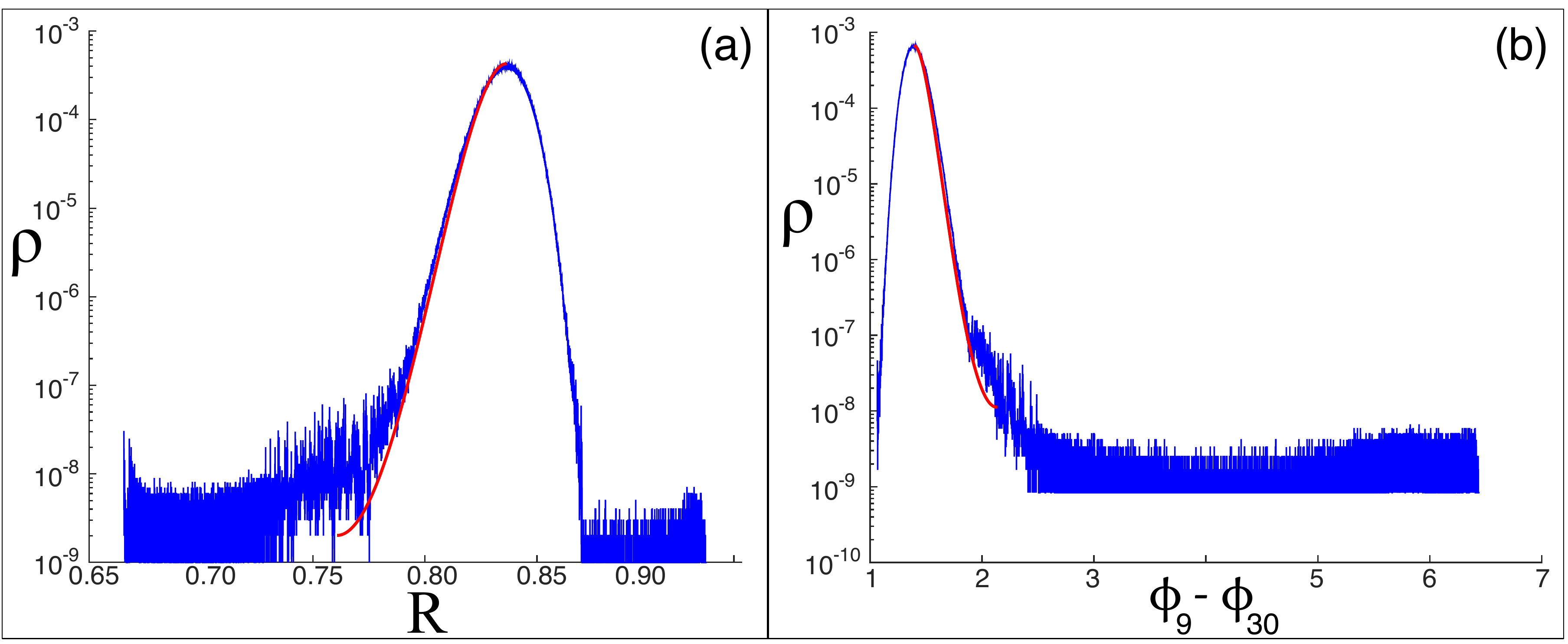}
\caption{Histograms of two network observables before a phase slip. (a) Coherence order parameter, $R\!=\!|\sum_{j}\!\exp\{i\phi_{j}\}\!/N|$ in a star-like network (Fig.\ref{StructureAndSlips} (b)): $J\!=\!0.718$, $D\!=\!0.001$, and $T_{c}\!=\!2.0$. (b) Phase difference between two oscillators in the IEEE network (Fig.\ref{StructureAndSlips} (d)): $J\!=\!0.325$, $D\!=\!0.001$, and $T_{c}\!=\!2.0$ \cite{F3}.}
\label{Distribution}
\end{figure}

First, we consider the action, Eq.(\ref{Action}), or probability exponent. Figure \ref{Distribution} shows histograms of two observables on logarithmic scale for two networks. The histograms were built from time-series data of ${\vec{\phi}}(t)$: starting from ${\vec{\phi}}^{*}$ at $t\!=\!0$ and ending at a time, $T$, when a phase slip occurred between any two connected oscillators in the networks. The histograms were averaged over ten simulations with different random number seeds, and are shown in blue. Predictions are shown in red for H1-- found by solving Eqs.(\ref{Action}-\ref{Lagrange3}) with boundary conditions \cite{HindesEPL,HindesSR2017}.    

\begin{figure}
 \centering
\includegraphics[scale=0.30]{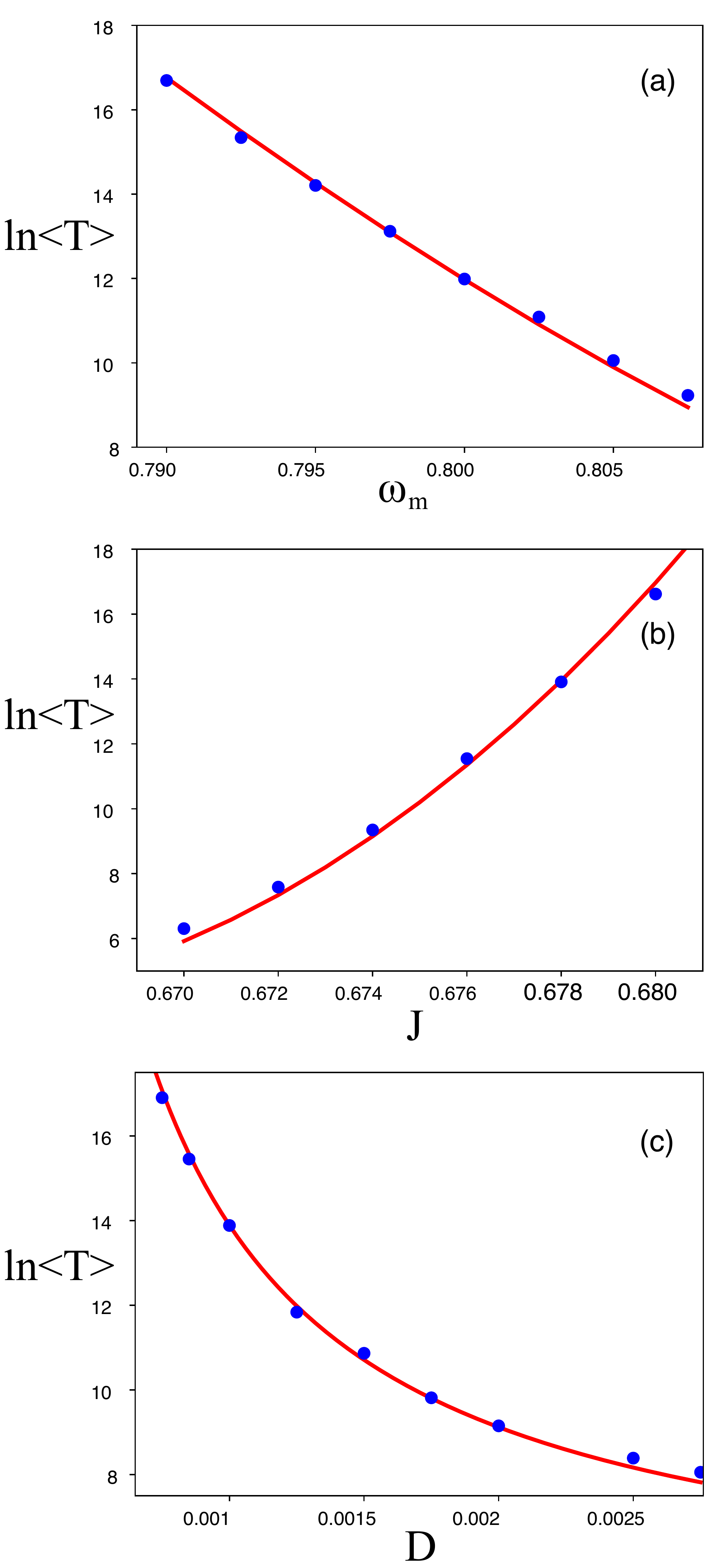}
\caption{Average phase-slip times versus network parameters: (a) natural frequency of the fastest oscillator, $\omega_{m}\!=\!\max_{i}\{\omega_{i}\}$, in a star-like network, Fig.\ref{StructureAndSlips} (b) ($J\!=\!0.718$, $D\!=\!0.001$, $T_{c}\!=\!2.0$). (b) coupling in the complete graph ($N\!=\!20$, $D\!=\!0.001$, $T_{c}\!=\!6.35$). (c) noise amplitude in a block network\cite{F3} ($J\!=\!0.1275$, $T_{c}\!=\!5.0$). The constant pre-factor in Eq.(\ref{Time}) was fitted for each network.}
\label{Times}
\end{figure}

Similarly, we can compare the average time scale over which slips occur, $\left<T\right>$, as a function of network parameters. Rare events are Poisson processes with rates proportional to their probabilities. Therefore, we expect   
\begin{align}
\label{Time}
\ln{\!\left<T\right>}=\mathcal{S}({\vec{\phi}}^{s})+\text{B}, 
\end{align}
where B is assumed to be an order one pre-factor\cite{Ira}. Example comparisons are given in Fig.\ref{Times}, where simulation averages are shown in blue and predictions in red with excellent agreement. Each panel demonstrates an intuitive result for $\ln{\!\left<T\right>}$: (a) dispersing the natural frequencies decreases the slip times, (b) increasing the coupling increases the slip times, and (c) increasing the noise amplitude decreases the slip times-- all exponentially.  

\section{\label{sec:Conclusion}CONCLUSION}
There is great interest in understanding how rare and extreme events occur in complex dynamical systems. One of the most broadly applicable classes of such dynamical systems is a network of nonlinear oscillators. In this work we have described analytically how noise drives phase-locked synchronized Kuramoto networks to saddle locked states, in the small-noise limit. Once at a saddle, an unstable mode dynamically separates groups of nodes-- resulting in phase slips. We have shown that different topologies and frequency distributions showcase different patterns and scaling-laws for slips and their probabilities. In particular, we showed that for sparse networks the slips occur between effectively disconnected subgraphs and the probability-exponent scales linearly with the critical coupling. In contrast for dense networks with continuous frequency distributions, the probability-exponent saturates to a constant value as the number of nodes becomes large. The latter occurs because individual oscillators on the edges of the frequency distribution fluctuate to saddles and slip. As a consequence, a network's coherence remains approximately constant in time during a noise-induced slip.     

Of course, the Kuramoto model is a simplified approximation of more general kinds of oscillator networks, such as networked limit-cycle oscillators with amplitude dynamics and networked power systems. Moreover in more realistic settings, networks may exhibit only partial synchronization, and noise may contain topological (e.g., spatial) correlations as well as intermittent and pulsed perturbations. Nevertheless, this work provides a foundation for analytically studying such issues, and others, related to rare events in nonlinear oscillator networks.   
\section{\label{sec:SM}SUPPLEMENTARY MATERIAL}
See supplementary material for network details, supporting calculations, and computational methods. 

\section*{\label{sec:Ack}ACKNOWLEDGMENTS}
\noindent JH is a National Research Council postdoctoral fellow. IBS was
supported by the U.S. Naval Research Laboratory funding (N0001414WX00023) and the Office of Naval Research (N0001416WX00657) and (N0001416WX01643).  

\end{document}